\documentclass[12pt]{article}
\usepackage{epsfig}
\usepackage{amssymb}
\usepackage{amsmath}
\usepackage{amsfonts}
\usepackage{graphicx}
\usepackage{mathrsfs}
\usepackage[dvips]{color}
\usepackage{multirow}

\newcommand{\bsigma}{\boldsymbol{\sigma}}

\newcommand{\R}{\mathbb{R}}
\newcommand{\C}{\mathbb{C}}
\newcommand{\Z}{\mathbb{Z}}

\newcommand{\fa}{\mathfrak{a}}
\newcommand{\fb}{\mathfrak{b}}

\newcommand{\ff}{\mathfrak{f}}

\newcommand{\fZ}{\mathfrak{Z}}

\newcommand{\bk}{\mathbf{k}}

\newcommand{\bfr}{\mathbf{r}}

\newcommand{\bs}{\mathbf{s}}

\newcommand{\bv}{\mathbf{v}}

\newcommand{\bI}{\mathbf{I}}

\newcommand{\bcL}{\boldsymbol{\cL}}
\newcommand{\bM}{\mathbf{M}}

\newcommand{\cB}{\mathcal{B}}

\newcommand{\cH}{\mathcal{H}}

\newcommand{\cF}{\mathcal{F}}

\newcommand{\cK}{\mathcal{K}}
\newcommand{\cL}{\mathcal{L}}

\newcommand{\cU}{\mathcal{U}}

\newcommand{\be}{\begin{equation}}
\newcommand{\ee}{\end{equation}}
\newcommand{\bea}{\begin{eqnarray}}
\newcommand{\eea}{\end{eqnarray}}
\newcommand{\nn}{\nonumber}
\newcommand{\kt}{\rangle}
\newcommand{\br}{\langle}

\newcommand{\ed}{\end{document}}

\newcommand{\bi}{\begin{itemize}}
\newcommand{\ei}{\end{itemize}}
\newcommand{\xto}{\Rightarrow}

\newcommand{\bce}{\begin{center}}
\newcommand{\ece}{\end{center}}
\newcommand{\sA}{\mathscr{A}}
\newcommand{\sB}{\mathscr{B}}
\newcommand{\sC}{\mathscr{C}}
\newcommand{\sD}{\mathscr{D}}

\newcommand{\sF}{\mathscr{F}}

\newcommand{\sH}{\mathscr{H}}

\newcommand{\sN}{\mathscr{N}}

\newcommand{\sR}{\mathscr{R}}

\newcommand{\sV}{\mathscr{V}}

\newcommand{\bcB}{{\boldsymbol{\cB}}}

\newcommand{\bcK}{{\boldsymbol{\cK}}}

\newcommand{\bcH}{{\boldsymbol{\cH}}}
\newcommand{\bcU}{{\boldsymbol{\cU}}}

\newcommand{\bvarpi}{{\boldsymbol{\varpi}}}

\newcommand{\bzero}{{\boldsymbol{0}}}

\newcommand{\for}{{\mbox{\rm for}}}
\newcommand{\dom}{{\mbox{\rm Dom}}}
\newcommand{\ran}{{\mbox{\rm Ran}}}

\newcommand{\bigimath}{{\mbox{\Large$\imath$}}}

\begin{document}

\title{Existence of the transfer matrix for a class of nonlocal potentials in two dimensions}


\author{Farhang Loran\thanks{E-mail address: loran@iut.ac.ir}~ and
Ali~Mostafazadeh\thanks{E-mail address:
amostafazadeh@ku.edu.tr}\\[6pt]
$^{*}$Department of Physics, Isfahan University of Technology, \\ Isfahan 84156-83111, Iran\\[6pt]
$^\dagger$Departments of Mathematics and Physics, Ko\c{c}
University,\\  34450 Sar{\i}yer, Istanbul, Turkey}

\date{ }
\maketitle

\begin{abstract}

Evanescent waves are waves that decay or grow exponentially in regions of the space void of interaction. In potential scattering defined by the Schr\"odinger equation, $(-\nabla^2+v)\psi=k^2\psi$ for a local potential $v$, they arise in dimensions greater than one and are present regardless of the details of $v$. The approximation in which one ignores the contributions of the evanescent waves to the scattering process corresponds to replacing $v$ with a certain energy-dependent nonlocal potential $\widehat{\mathscr{V}}_k$. We present a dynamical formulation of the stationary scattering for $\widehat{\mathscr{V}}_k$ in two dimensions, where the scattering data are related to the dynamics of a quantum system having a non-self-adjoint, unbounded, and nonstationary Hamiltonian operator. The evolution operator for this system determines a two-dimensional analog of the transfer matrix of stationary scattering in one dimension which contains the information about the scattering properties of the potential. Under rather general conditions on $v$, we establish the strong convergence of the Dyson series expansion of the evolution operator and prove the existence of the transfer matrix for $\widehat{\mathscr{V}}_k$ as a densely-defined operator acting in $\C^2\otimes L^2(-k,k)$. 



\end{abstract}

\section{Introduction}
\label{S1}

Recently we have proposed a formulation of stationary scattering in two and three dimensions that is based on a multi-dimensional generalization of the transfer matrix of potential scattering in one dimension \cite{pra-2021}. The earlier attempts in this direction \cite{pendry-1984,pendry-1990a,pendry-1990b,pendry-1996}  involved slicing the space along the scattering axis and discretizing the transverse degrees of freedom. This led to large numerical transfer matrices which allowed for a numerical treatment of the scattering problem. The approach pursued in \cite{pra-2021} is in sharp contrast, for it identifies the transfer matrix with a fundamental mathematical construct that is given by the time-evolution operator for an effective quantum system. Because the Hamiltonian operator for this system is an unbounded non-self-adjoint operator, the developments reported in \cite{pra-2021} are generally formal. In the present article we take a first step towards providing a mathematically rigorous basis for these developments. Specifically, we offer a comprehensive analysis of various operators entering the definition of the transfer matrix and give a proof of its existence for a class of energy-dependent nonlocal potentials in two dimensions. These potentials arise in an approximate scheme that involves neglecting the contribution of the evanescent waves to the scattering data \cite{p158}. 

Consider the scattering problem defined by the stationary Schr\"odinger equation,
	\be
	[-\partial_x^2-\partial_y^2+v(x,y)]\psi(x,y)=k^2\psi(x,y),~~~~~(x,y)\in\R^2,
	\label{sch-eq}
	\ee
where $v:\R^2\to\C$ is a real or complex short-range potential \cite{yafaev}, $k$ is a wavenumber, and we have adopted units where $\hbar^2/2m=1$. Performing a partial Fourier transformation of both sides of (\ref{sch-eq}) with respect to $y$, we find 
	\be
	-\partial_x^2\tilde\psi(x,p)+v(x,i\partial_p)\tilde\psi(x,p)=\varpi(p)^2\tilde\psi(x,p),~~~~~~~~~x,p\in\R,
	\label{sch-eq-171}	
	\ee
where $\tilde\psi(x,p)$ is the Fourier transform of $\psi(x,y)$ with respect to $y$, i.e.,
	\be
	\tilde\psi(x,p):=\int_{-\infty}^\infty dy\:e^{-ipy}\psi(x,y),
	\label{Fouier-trans-171}
	\ee
and we have introduced
	\bea
	&&v(x,i\partial_p) f(p):=\frac{1}{2\pi}\int_{-\infty}^\infty dq\:\tilde v(x,p-q) f(q),
	\label{hat-v1}\\
	&&\varpi(p):=\left\{\begin{array}{ccc}
	\sqrt{k^2-p^2} & \for & |p|< k,\\
	i\sqrt{p^2-k^2} & \for & |p|\geq k.\end{array}\right.
	\label{varpi}
	\eea
In view of (\ref{Fouier-trans-171}), we can express $\psi$ in the form  $\psi=\psi_{\rm os}+\psi_{\rm ev}$, where
	\bea
	\psi_{\rm os}(x,y)&:=&\frac{1}{2\pi}\int_{-k}^k dp\:e^{ipy}\tilde\psi(x,p),
	\label{os-wave}\\
	\psi_{\rm ev}(x,y)&:=&\frac{1}{2\pi}\left[\int_{-\infty}^{-k} dp\:e^{ipy}\tilde\psi(x,p)+
	\int_{k}^{\infty} dp\:e^{ipy}\tilde\psi(x,p)\right].
	\label{ev-wave}
	\eea
	
Suppose that there is some $\Omega\subseteq\R^2$ such that $v(x,y)=0$ for $(x,y)\in\Omega$. Then the general solution of (\ref{sch-eq}) in $\Omega$ takes the form,
	\be
	\psi(x,y)=\int_{-\infty}^\infty \frac{dp}{4\pi^2\varpi(p)}\Big[
	\sA(p) e^{i\varpi(p)x}+\sB(p) e^{-i\varpi(p)x}\Big]e^{ip y},
	\label{eqqq8}
	\ee
where $\sA,\sB:\R\to\C$ are some coefficient functions. This in turn implies that  
	\bea
	\psi_{\rm os}(x,y)&=&\int_{-k}^k \frac{dp}{4\pi^2\varpi(p)}
	\Big[A(p) e^{i\varpi(p)x}+B(p) e^{-i\varpi(p)x}\Big]e^{ipy},
	\label{os-wave-free}\\
	\psi_{\rm ev}(x,y)&=&\int_{-\infty}^{-k} 
	\frac{dp}{4\pi^2\varpi(p)}
	\Big[\sA(p)e^{-|\varpi(p)|x}+\sB(p)e^{|\varpi(p)|x}\Big]e^{ipy}+\nn\\
	&&\int_{k}^{\infty} \frac{dp}{4\pi^2\varpi(p)}
	\Big[\sA(p)e^{-|\varpi(p)|x}+\sB(p)e^{|\varpi(p)|x}\Big]e^{ipy},
	\label{ev-wave-free}
	\eea
where $(x,y)\in\Omega$ and 
	\begin{align}
	&A(p):=\left\{\begin{array}{ccc}
	\sA(p)&\for& |p|<k,\\
	0&\for& |p|\geq k,\end{array}\right.
	&&B(p):=\left\{\begin{array}{ccc}
	\sB(p)&\for& |p|<k,\\
	0&\for& |p|\geq k.\end{array}\right.
	\label{AB=}
	\end{align}
According to (\ref{os-wave-free}) and (\ref{ev-wave-free}),  $\psi_{\rm os}$ is a superposition of the plane-wave solutions of (\ref{sch-eq}) which are oscillating functions of $x$, whereas $\psi_{\rm ev}$ is the superposition of exponentially growing or decaying functions of $x$. We therefore call $\psi_{\rm os}$ and $\psi_{\rm ev}$ the oscillating and evanescent waves, respectively. 

Let $\sF$ denote the vector space of functions (tempered distributions) of $p$ that possess Fourier transform, and $\sF_k$ be the subspace of $\sF$ consisting of elements whose support lies in the interval $(-k,k)$. Then $A,B\in\sF_k$.

If $v$ is a short-range potential \cite{yafaev}, the bounded solutions of (\ref{sch-eq}) tend to the superposition of the plane-wave solutions as $x\to\pm\infty$, i.e., there are  $A_\pm,B_\pm\in\sF_k$ such that
	\be
	\psi(x,y)\to\int_{-k}^k \frac{dp}{4\pi^2\varpi(p)}
	\Big[A_\pm(p) e^{i\varpi(p)x}+B_\pm(p) e^{-i\varpi(p)x}\Big]e^{ipy}~~~~\for~~~~x\to\pm\infty.
	\label{asym}
	\ee
Ref.~\cite{pra-2021} identifies the fundamental transfer matrix for the potential $v$ with a $2\times 2$ matrix $\widehat\bM$ with operator entries $\widehat M_{ij}$ that satisfies\footnote{The coefficient functions $A_\pm$ and $B_\pm$ correspond to those denoted by  $\breve A_\pm$ and $\breve B_\pm$ in Ref.~\cite{pra-2021}.}
	\be
	\widehat\bM\left[\begin{array}{c}
	A_-\\
	B_-\end{array}\right]=
	\left[\begin{array}{c}
	A_+\\
	B_+\end{array}\right].
	\label{M-def}
	\ee
Because $A_\pm,B_\pm\in\sF_k$,  $\widehat M_{ij}$ and $\widehat\bM$ are respectively linear operators acting in $\sF_k$ and
	\be
	\sF_k^{2\times 1}:=\C^2\otimes\sF_k:=\left\{\:\left[\begin{array}{c}
	\phi_+\\
	\phi_-\end{array}\right]~\Big|~\phi_\pm\in\sF_k\:\right\}.\nn
	\ee
Eq.~(\ref{M-def}) coincides with the defining relation for the transfer matrix in one dimension except that in one dimension $A_\pm$ and $B_\pm$ are complex numbers and the transfer matrix is a numerical matrix \cite{prl-2009,sanchez-soto}. Another common feature of the transfer matrix (\ref{M-def}) and its one-dimensional analog is that it stores the information about the scattering properties of the potential \cite{pra-2021}. We provide a brief description of the relationship between the transfer matrix and the scattering amplitude of the potential in the appendix.

Next, suppose that the $f(q)$ appearing on the right-hand side of (\ref{hat-v1}) vanishes for all $|q|\geq k$, and that $v\neq 0$. Then it is not difficult to show that the left-hand side of (\ref{hat-v1}) does not vanish for all $|p|\geq k$, \cite{p158}. Making use of this observation in (\ref{sch-eq-171}), we infer that $v$ always couples to the evanescent waves, and $\psi_{\rm os}$ can never solve the Schr\"odinger equation~(\ref{sch-eq}) for $v\neq 0$. It is also clear from (\ref{os-wave}) that the requirement, $\psi=\psi_{\rm os}$, implies $\tilde\psi(x,p)=0$ for $|p|\geq 0$. Such a function satisfies an equation of the form (\ref{sch-eq-171}), if we replace $v(x,i\partial_p)$ with the operator $\widehat V_k(x)$ acting in $\sF$ according to
	\be
	\big(\widehat V_k(x) f\big)(p):=
	\frac{\chi_k(p)}{2\pi}\int_{-k}^k dq\:\tilde v(x,p-q) f(q),
	\ee
where
	\[\chi_k(p):=\left\{\begin{array}{ccc}
	1 &\for & |p|<k,\\
	0 &\for & |p|\geq k.\end{array}\right.\]
This corresponds to replacing $v$ in (\ref{sch-eq}) with the energy-dependent nonlocal potential given by
	\be
	(\widehat\sV_k\psi)(x,y):=\frac{1}{4\pi^2}\int_{-k}^k dp
	\int_{-k}^k dq\: e^{ipy}\:\tilde v(x,p-q)\tilde\psi(x,q).
	\label{nonlocal-v-2d-1}
	\ee 
This relation defines an operator acting in $L^2(\R^2)$ which satisfies
	\be
	\widehat\sV_k=\widehat\Pi_k v(\widehat x,\widehat y)\widehat\Pi_k.
	\label{nonlocal-v-2d-2}
	\ee
Here $\widehat\Pi_k$ is the projection operators given by,
	\be
	(\widehat\Pi_k\psi)(x,y):=\frac{1}{{2\pi}}\int_{-k}^k dp\: e^{ipy}\tilde\psi(x,p),
	\label{Pi-k}
	\ee
and $\widehat x$ and $\widehat y$ are respectively the multiplication (position) operators: $(\widehat x\psi)(x,y):=x\psi(x,y)$ and $(\widehat y\psi)(x,y):=y\psi(x,y)$.

The right-hand side of (\ref{asym}) is a superposition of oscillating waves. This suggests that although the solutions of the Schr\"odinger equation~(\ref{sch-eq}) for $v\neq 0$ are different from the ones for the nonlocal potential $\widehat\sV_k$, their asymptotic form need not differ appreciably. This means that the solution to the scattering problem for $\widehat \sV_k$ may provide a reliable approximation for the solution to the scattering problem defined by $v$ at the energy $k^2$. Ref.~\cite{p158} offers evidence for the exactness of this approximation for potentials $v$ satisfying $\tilde v(x,p)=0$ for $p\leq 0$ or $p\geq 0$. This provides our basic motivation for the study of the scattering properties of the nonlocal potentials $\widehat\sV_k$.

The transfer matrix $\widehat\bM$ for the nonlocal potential $\widehat\sV_k$ turns out to be given by the following formal Dyson series \cite{p158}.
	\be
	\widehat\bI+\sum_{n=1}^\infty (-i)^n
         \int_{-\infty}^{\infty} \!\!dx_n\int_{-\infty}^{x_n} \!\!dx_{n-1}
         \cdots\int_{-\infty}^{x_2} \!\!dx_1\,
         \widehat{\bcH}(x_n)\widehat{\bcH}(x_{n-1})\cdots\widehat{\bcH}(x_1),
         \label{Dyson-bcU-infinite}
         \ee
where $\widehat\bI$ is the identity operator acting in $\sF_k^{2\times 1}$, $\widehat\bcH(x)$ is an effective Hamiltonian operator given by
	\bea
   	&&\widehat\bcH(x):=\frac{1}{2}\,e^{-i\widehat\varpi x\bsigma_3}\,
         \widehat V_k(x)\,\widehat\varpi ^{-1}\bcK\, e^{i\widehat\varpi x\bsigma_3},
         \label{bcH-def}\\[3pt]
         &&(\widehat\varpi f)(p):=\varpi(p)f(p),~~~~~~(\widehat p f)(p):=p f(p),\\[3pt]
	&&\bcK:=\left[\begin{array}{cc}
    	1 & 1 \\
    	-1 & -1\end{array}\right]=\bsigma_3+i\bsigma_2,
    	\label{bcK-def}
    	\eea
$f\in\sF_k$, and $\bsigma_j$ are the Pauli matrices;
	\begin{align}
	&\bsigma_1:=\left[\begin{array}{cc}
	0 & 1\\
	1 & 0\end{array}\right],
	&& \bsigma_2:=\left[\begin{array}{cc}
	0 & -i\\
	i & 0\end{array}\right],
	&&\bsigma_3:=\left[\begin{array}{cc}
	1 & 0\\
	0 & -1\end{array}\right].
	\label{Pauli}
	\end{align}
The purpose of this article is to make the above crude description of the transfer matrix $\widehat\bM$, Hamiltonian operator $\widehat\bcH(x)$, and the dynamics it generates into mathematically rigorous statements. 

The outline of this article is as follows. In Sec.~2, we collect a few basic mathematical results about the analogs of the nonlocal potential $\widehat\sV_k$ in one dimension. In Sec.~3, we extend these to two dimensions by confining our attention to a particular class of potentials $v$ for which our analysis apply. In Sec.~4 we outline the dynamical formulation of the scattering problem for $\widehat\sV_k$. Here we identify $\widehat\bcH(x)$ with a densely-defined linear operator acting in the Hilbert space,
	\be
	\sH:=\C^2\otimes L^2(-k,k)=\left\{\:\left[\begin{array}{c}
	\xi_+\\
	\xi_-\end{array}\right]~\Big|~\xi_\pm\in L^2(-k,k)\:\right\},
	\label{sH-def}
	\ee
establish the $x$-independence of the domain of $\widehat\bcH(x)$, and show that the range of $\widehat\bcH(x)$ lies in its domain. In Sec.~5, we establish the strong convergence of the Dyson series, 
	\[\widehat\bI+\sum_{n=1}^\infty (-i)^n
         \int_{x_0}^{x} \!\!dx_n\int_{x_0}^{x_n} \!\!dx_{n-1}
         \cdots\int_{x_0}^{x_2} \!\!dx_1\,
         \widehat{\bcH}(x_n)\widehat{\bcH}(x_{n-1})\cdots\widehat{\bcH}(x_1),\]
for $x_0,x\in\R$ such that $x_0\leq x$. This gives a densely-defined linear operator $\widehat\bcU(x,x_0)$ that serves as the evolution operator for the Hamiltonian operator $\widehat\bcH(x)$. Finally, we prove the existence of the strong limit of $\widehat\bcU(x,x_0)$ as $x_0\to-\infty$ and $x\to+\infty$, which we identify with the transfer matrix of $\widehat\sV_k$.

\section{Analogs of nonlocal potentials $\widehat\sV_k$ in one dimension}
\label{S4-1}

Consider an integrable potential, $v:\R\to\C$, in one dimension, i.e., $v\in L^1(\R)$. Let $\widehat\Pi_k$ be the projection operator defined on $L^2(\R)$ by (\ref{Pi-k}), and $\widehat\sV_k,\widehat{V}_k:L^2(\R)\to L^2(\R)$ be linear operators given by 
	\begin{align}
	&\widehat\sV_k:=\widehat\Pi_k\, v(\widehat y)\, \widehat\Pi_k,
	&& \widehat{V}_k:=\cF\,\widehat\sV_k\,\cF^{-1},
	\label{hat-v}
	\end{align}
where $\widehat y: L^2(\R)\to L^2(\R)$ is the multiplication (position) operator: $(\widehat y\phi)(y):=y\phi(y)$, and $\cF,\cF^{-1}:L^2(\R)\to L^2(\R)$ respectively label the Fourier transformation and its inverse;
	\begin{align}
	&(\cF\phi)(p):=\tilde\phi(p):=\int_{-\infty}^\infty dy\, e^{-ipy}\phi(y),
	&&(\cF^{-1}\tilde\phi)(y):=\frac{1}{2\pi}\int_{-\infty}^\infty dp\, e^{ipy}
	\tilde\phi(p).
	\label{Fourier}
	\end{align} 
Then, according to (\ref{Pi-k}) and (\ref{hat-v}), for every function $\phi$ in the domain of $\widehat\sV_k$ and all $y\in\R$,
	\bea
	(\widehat\sV_k\phi)(y)&=&\frac{1}{4\pi^2}\int_{-k}^k dp
	\int_{-k}^k dq\: e^{ipy}\:\tilde v(p-q)\tilde\phi(q).
	\label{vk-y}
	\eea
In view of (\ref{hat-v}) -- (\ref{vk-y}), 
	\bea
	\big(\widehat{V}_k\tilde\phi\big)(p)=
	\big(\cF\,\widehat\sV_k \cF^{-1}\tilde\phi\big)(p)=
	\big(\cF\,\widehat\sV_k\phi\big)(p)=
	\frac{\chi_k(p)}{2\pi}\int_{-k}^k dq\, \tilde v(p-q)\tilde\phi(q).
	\label{id-1}
	\eea
	
Because $v\in L^1(\R)$, $\tilde v$ is uniformely continuous \cite{katznelson}. This in particular implies that $\tilde v$ is bounded on $[-2k,2k]$, i.e., there is $\mu\in\R^+$ such that for all $p\in[-2k,2k]$, $|\tilde v(p)|\leq\mu$. This together with the Cauchy-Schwarz inequality show that, for all $\phi\in L^2(\R)$ and $p\in[-k,k]$,
	\be
	\left|\int_{-k}^kdq\,\tilde v(p-q)\tilde\phi(q)\right|^2\leq 
	\left(\int_{-k}^kdq\,|\tilde v(p-q)|^2\right)
	\left(\int_{-k}^kdq\,|\tilde \phi(q)|^2\right)
	\leq 2k\mu^2\parallel\tilde\phi\parallel^2.
	\label{ineq-1}
	\ee
(\ref{hat-v}), (\ref{id-1}), and (\ref{ineq-1}) imply that $\widehat{V}_k$ and consequently $\widehat\sV_k$ are defined everywhere in $L^2(\R)$;
	\be
	\dom(\widehat\sV_k)=\dom(\widehat{V}_k)=L^2(\R).\nn
	\ee
{\bf Theorem~1:} $\widehat\sV_k$ and $\widehat{V}_k$ are normal bounded operators acting in $L^2(\R)$.\\[6pt]
{\bf Proof:} According to (\ref{hat-v}), (\ref{id-1}), (\ref{ineq-1}), and the fact that $\cF$ is a unitary operator,
	\bea
	\parallel \widehat\sV_k\phi\parallel^2=
	\parallel \widehat{V}_k\tilde\phi\parallel^2=\frac{1}{4\pi^2}
	\int_{-k}^kdp
	\left|\int_{-k}^kdq\,\tilde v(p-q)\tilde\phi(q)\right|^2\leq
	\left(\frac{k\mu}{\pi}\right)^{\!2}\!\!\parallel\tilde\phi\parallel^2=
	\left(\frac{k\mu}{\pi}\right)^{\!2}\!\!\parallel\phi\parallel^2.
	\eea
This shows that $\widehat\sV_k$ and $\widehat{V}_k$ are bounded. Next, we use (\ref{vk-y}) to deduce
	\be
	\widehat\sV_k^\dagger
	=\widehat\Pi_k^\dagger v(\widehat y)^\dagger \widehat\Pi_k^\dagger
	=\widehat\Pi_k v^*(\widehat y) \widehat\Pi_k,
	\label{v-k-dagger}
	\ee 
where $v^*:\R\to\C$ is the complex-conjugate of $v$, i.e., for all $y\in\R$, $v^*(y):=v(y)^*$. Because $v(\widehat y)$ and $v^*(\widehat y)$ commute, so do $\widehat\sV_k$ and $\widehat\sV_k^\dagger$. Hence $\widehat\sV_k$ is a normal operator. The same holds for $\widehat{V}_k$ by virtue of (\ref{hat-v}) and the fact that $\cF$ is a unitary operator.~~$\square$\vspace{6pt}

Consider the embedding $\bigimath:L^2(-k,k)\to L^2(\R)$ given by
	\be
	(\bigimath\,\xi)(p):=
	\left\{\begin{array}{ccc}
	\xi(p)&{\rm for}& p\in(-k,k),\\
	0 &{\rm for}& p\notin(-k,k),\end{array}\right.
	\label{embed}
	\ee
for all $\xi\in L^2(-k,k)$ and $p\in\R$. Clearly, $\bigimath$ is an isometry, and
	\[\dom(\bigimath^{-1})=
	\ran(\bigimath)=\left\{\psi\in L^2(\R)~|~\psi(p)=0~{\rm for}~|p|\geq k\right\}.\]
In view of (\ref{id-1}),  $\widehat{V}_k\bigimath$ maps elements of $L^2(-k,k)$ to $\ran(\bigimath)$. Therefore, 
	\be
	\widehat v_k:=\bigimath^{-1}\,\widehat{V}_k\;\bigimath,
	\label{Big-V}
	\ee
defines a linear operator acting in $L^2(-k,k)$. Because $\bigimath$ is an isometry and $\widehat{V}_k$ is a bounded operator with domain $L^2(\R)$, $\widehat v_k$ is a bounded operator with domain $L^2(-k,k)$. We can use (\ref{id-1}), (\ref{embed}), and (\ref{Big-V}) to show that for all $\xi\in L^2(-k,k)$ and all $p\in(-k,k)$,
	\be
	(\widehat v_k\xi)(p)=\frac{1}{2\pi}\int_{-k}^k dq\, \tilde v(p-q)\xi(q).
	\ee
Because $\tilde v$ is continuous, $\widehat v_k:L^2(-k,k)\to L^2(-k,k)$ is a Hilbert-Schmidt operator \cite{Reed-Simon1}. In particular, it is compact. Because $\widehat{V}_k$ is a normal bounded operator, (\ref{Big-V}) shows that the same holds for $\widehat v_k$. This proves the following theorem.\\[6pt]
\noindent{\bf Theorem~2:} $\widehat v_k$ is a normal Hilbert-Schmidt operator acting in $L^2(-k,k)$.\\[6pt]
\noindent An immediate consequence of this theorem is the existence of an orthonormal basis of $L^2(-k,k)$ consisting of the eigenvectors $\psi_n$ of $\widehat v_k$ with possibly repeated eigenvalues $\nu_n$ so that \cite{kato,Beauzamy}
	\be
	\widehat v_k=
	\sum_{n=0}^\infty\nu_n\,\br\psi_n|\cdot\kt\,\psi_n.
	\label{V-expand}
	\ee
Furthermore, because $\widehat v_k$ is a normal compact operator, there is $n_\star\in\Z^+$ such that
	\be
	\parallel \widehat v_k\parallel_0=|\nu_{n_\star}|,
	\label{V-norm=}
	\ee
where $\parallel\cdot\parallel_0$ stands for the operator norm \cite{kato};
	\be
	\parallel\widehat v_k\parallel_0:=\mathop{\mbox{sup}}\limits_{\xi\neq 0}\: \frac{\parallel \widehat v_k\xi\parallel}{\parallel\xi\parallel}.
	\ee

\section{Generalization to two dimensions}

\noindent {\bf Definition}: Let $s,k\in\R^+$. Then a function $v:\R^2\to\C$ is said to belong to the class $\sC_{s,k}$ if it fulfills the following conditions.
	\begin{itemize}
	\item[]C1: For all $x\in\R$, $v(x,\cdot):\R\to\C$ is integrable, i.e., $\int_{-\infty}^\infty dy\:|v(x,y)|<\infty$, so that its Fourier transform $\tilde v(x,\cdot):\R\to\C$ exists and is uniformly continuous.
	
	\item[]C2: The function $\mathring v_k:\R \to L^\infty(-2k,2k)$ defined by
		\be
		\big(\mathring v_k(x)\big)(p):=\tilde v(x,p),~~~~~x\in \R,~p\in(-2k,2k),
		\ee
	is piecewise continuous, i.e., for every closed interval $I$ in $\R$, there are finitely many open subintervals, $I_1,I_2,\cdots,I_j$, such that $I$ is the closure of $\bigcup_{i=1}^j I_i$ and $\mathring v_k$ is continuous on $I_i$ for all $i\in\{1,2,\cdots,j\}$. The latter requirement means that for all $x,x'\in I_i$ and all $\epsilon\in\R^+$, there is some $\delta\in\R^+$ such that
		\be
		|x-x'|<\delta~~\xto~\mathop{\mbox{sup}}\limits_{{p\in(-2k,2k)}} 
		|\tilde v(x,p)-\tilde v(x',p)|<\epsilon.
		\label{C2-condi}
		\ee
	
	\item[]C3: $v$ is bounded, and there are $\alpha,\beta,\sigma\in\R^+$ such that $\sigma>s$ and for all $(x,y)\in\R^2$,
		\be
		|v(x,y)|\:\leq\:\displaystyle\frac{\beta}{(1+|x|)^\sigma}~~~\for~~~|x|\geq\alpha.
		\label{v-bound}
		\ee
	\end{itemize}
Condition C1 allows us to apply the constructions of Sec.~\ref{S4-1} by letting $v(x,\cdot)$ play the role of the potential $v$ of Sec.~\ref{S4-1}. In particular, we can use $v(x,\cdot)$ to introduce an associated one-parameter family of normal Hilbert-Schmidt operators acting in $L^2(-k,k)$. We view this as an operator function $\widehat v_k$ that maps $\R$ to the space of normal Hilbert-Schmidt operators acting in $L^2(-k,k)$; for each $x\in\R$, $\xi\in L^2(-k,k)$, and $p\in(-k,k)$,
	\be
	\big(\widehat v_k(x)\xi\big)(p):=
	\frac{1}{2\pi}\int_{-k}^k dq\, \tilde v(x,p-q)\xi(q).
	\label{v-hat-def}
	\ee
As we show below, C2 implies that $\parallel\widehat v_k(\cdot)\parallel_0$ is a piecewise continuous function, while C3 puts an upper bound on $\parallel\widehat v_k(x)\parallel_0$ for $|x|\geq\alpha$.\vspace{6pt}

\noindent {\bf Lemma~1:} Let $k\in\R^+$ and $v:\R^2\to\C$ be a function that satisfies conditions C1 and C2. Then $\parallel\widehat v_k(\cdot)\parallel_0:\R\to [0,\infty)$ is a piecewise continuous function.\\[6pt]
\noindent {\bf Proof:} Let $I$ be any closed interval in $\R$. According to C2 there are open subintervals, $I_1,I_2,\cdots,I_j$, such that $I$ is the closure of $\bigcup_{i=1}^j I_i$ and for all $i\in\{1,2,\cdots,j\}$, $\mathring v_k$ is continuous on $I_i$. Let $i\in\{1,2,\cdots,j\}$ and $\epsilon\in\R^+$ be arbitrary, and $\epsilon':=\pi \epsilon/k$. Because $\mathring v_k$ is continuous on $I_i$ and $\epsilon'\in\R^+$, for all $x,x'\in I_i$, there is some $\delta\in\R^+$ such 
		\be
		|x-x'|<\delta~~\xto~\mathop{\mbox{sup}}\limits_{{p\in(-2k,2k)}} 
		|\tilde v(x,p)-\tilde v(x',p)|<\epsilon'.
		\label{limit-1}
		\ee
Let $\phi\in L^2(-k,k)\setminus\{0\}$ be arbitrary, and suppose that $x,x'\in I_i$ and $|x-x'|<\delta$. Then, (\ref{v-hat-def}) and (\ref{limit-1}) imply
	\bea
	\Big|\big([\widehat v_k(x)-\widehat v_k(x')]\phi\big)(p)\Big|&=&
	\frac{1}{2\pi}\left|\int_{-k}^k dq[\tilde v(x,p-q)-\tilde v(x',p-q)]\phi(q)\right|\nn\\
	&\leq&\frac{1}{2\pi}\int_{-k}^k dq\: |\tilde v(x,p-q)-\tilde v(x',p-q)|\; |\phi(q)|\nn\\
	&\leq&\frac{1}{2\pi}\mathop{\mbox{sup}}\limits_{{p\in(-2k,2k)}} 
	|\tilde v(x,p)-\tilde v(x',p)|\int_{-k}^k dq\: |\phi(q)|\nn\\
	&<& \frac{\epsilon'}{\pi}\sqrt{\frac{k}{2}}\,\parallel\phi\parallel
	\,=\frac{\epsilon \parallel\phi\parallel}{\sqrt{2k}},
	\label{limit-2}
	\eea
where we have made use of the fact that 
	$\int_{-k}^k dq\: |\phi(q)|\leq\sqrt{\int_{-k}^k dq}\;\sqrt{
	\int_{-k}^k dq\: |\phi(q)|^2}=\sqrt{2k}
	\parallel \phi\parallel$. 
The following is a simple consequence of (\ref{limit-2}).
	\be
	\frac{\parallel [\widehat v_k(x)-\widehat v_k(x')]\phi\parallel}{\parallel\phi\parallel}
	=\frac{1}{2\pi \parallel\phi\parallel}\sqrt{\int_{-k}^k dp
	\Big|\big([\widehat v_k(x)-\widehat v_k(x')]\phi\big)(p)\Big|^2}<\epsilon.\nn	
	\ee
Because this relation holds for all $\phi\in L^2(-k,k)\setminus\{0\}$, it implies $\parallel \widehat v_k(x)-\widehat v_k(x')\parallel_0<\epsilon$. This completes the proof that for all $x,x'\in I_i$, $|x-x'|<\delta$ implies $\parallel \widehat v_k(x)-\widehat v_k(x')\parallel_0<\epsilon$. Hence, $\parallel \widehat v_k(\cdot)\parallel_0$ is continuous on $I_i$. This together with the fact that $I$ is an arbitrary closed subset of $\R$ and the closure of $\bigcup_{i=1}^j I_i$ coincides with $I$ show that $\parallel \widehat v_k(\cdot)\parallel_0$ is piecewise continuous.~~~$\square$\vspace{6pt}
	
\noindent {\bf Lemma~2:} Let $k,s\in\R^+$ and $v\in\sC_{s,k}$\,, so that (\ref{v-bound}) holds for some $\alpha,\beta,\sigma\in\R^+$ with $\sigma>s$. Then, 
	\be
	\parallel\widehat v_k(x)\parallel_0\:\leq\:\frac{2\pi\beta}{(1+|x|)^{\sigma}}~~\for~~|x|\geq\alpha.
	\label{lemma1}
	\ee
\noindent {\bf Proof:} Because for each $x\in\R$, $\widehat v_k(x)$ is a compact normal operator, it has an eigenvalue $\nu_{\star}(x)$ such that $\parallel\widehat v_k(x)\parallel_0=|\nu_*(x)|$, \cite{kato}. Therefore, to prove (\ref{lemma1}), it suffices to show that
	\be
	|\nu_*(x)|\:\leq\:
	\frac{2\pi\beta}{(1+|x|)^{\sigma}}~~\for~~|x|\geq\alpha.
	\label{Lemma1-1}
	\ee
Let $\xi_\star(x)\in L^2(-k,k)$ be an eigenvector of $\widehat v_k(x)$ with eigenvalue $\nu_\star(x)$, and  introduce 
	\begin{align}
	&\phi_\star(x):=\bigimath\,\xi_\star(x),
	&&\breve\phi_\star(x):=\cF^{-1}\phi_\star(x),
	&&\breve\phi_\star(x,y):=\big(\breve\phi_\star(x)\big)(y),\nn\\
	&\xi_\star(x,p):=\big(\xi_\star(x)\big)(p), 
	&&\phi_\star(x,q):=\big(\phi_\star(x)\big)(q),\nn
	\end{align}
where $\bigimath$ is the embedding of $L^2(-k,k)$ into $L^2(\R)$ given by (\ref{embed}), $p\in(-k,k)$, $q\in\R$, and $y\in\R$. Then,
	\bea
	\breve\phi_\star(x,y)&=&\frac{1}{2\pi}\int_{-\infty}^\infty dp\: e^{ipy}\:\phi_\star(x,p)=
	\frac{1}{2\pi}\int_{-k}^k dp\: e^{ipy}\xi_\star(x,p),
	\label{lemma1-2}\\
	\parallel\xi_\star(x)\parallel&=&\parallel\phi_\star(x)\parallel~=~\parallel\breve\phi_\star(x)\parallel,
	\label{norm=norm}
	\eea
and
	\bea
	\nu_\star(x)\xi_\star(x,p)&=&\big(\widehat v_k(x)\xi_\star(x)\big)(p)
	=\frac{1}{2\pi}\int_{-k}^k dq\:\widetilde v(x,p-q)\xi_\star(x,q)\nn\\
	&=&\frac{1}{2\pi}\int_{-k}^k dq
	\int_{-\infty}^\infty\!\!dy\: e^{-iy(p-q)}v(x,y)\xi_\star(x,q)=
	\int_{-\infty}^\infty dy\:e^{-ipy}v(x,y)\breve\phi_\star(x,y).
	\nn
	\eea
Multiplying both sides of the latter equation by $\xi_\star(x,p)^*$, integrating over $p$, and using (\ref{v-bound}) and (\ref{lemma1-2}), we have
	\bea
	\left|\nu_\star(x)\right|\parallel\xi_\star(x)\parallel^2&=&
	\left|\int_{-k}^k\!\!dp
	\left[\xi_\star(x,p)^*\int_{-\infty}^\infty\!\! dy\:e^{-ipy}v(x,y)\breve\phi_\star(x,y)\right]\right|\nn\\
	&=&2\pi\left|\int_{-\infty}^\infty dy\:v(x,y)|\breve\phi_\star(x,y)|^2\right|
	\leq 2\pi\int_{-\infty}^\infty dy\:|v(x,y)||\breve\phi_\star(x,y)|^2\nn\\
	&\leq&\frac{2\pi\beta \parallel\breve\phi_\star(x)\parallel^2}{(1+|x|)^{\sigma}}~~~\for~~~|x|\geq\alpha.
	\eea
Because $\parallel\breve\phi_\star(x)\parallel=\parallel\xi_\star(x)\parallel\neq 0$, this implies (\ref{Lemma1-1}).~~~$\square$\vspace{6pt}

\section{Dynamical formulation of stationary scattering for $\widehat\sV_k$}
		
The scattering problem for the nonlocal potential $\widehat\sV_k$ involves bounded solutions of the Schr\"odinger equation, 
	\be
	(-\partial_x^2-\partial_y^2)\psi(x,y)+\big(\widehat\sV_k\psi\big)(x,y)=k^2\psi(x,y),~~~~~(x,y)\in\R^2.
	\label{sch-eq-nonlocal}
	\ee
We can express this in the form
	\be
	-\partial_x^2\tilde\psi(x,p)+
	\frac{\chi_k(p)}{2\pi}\int_{-k}^k dq\:
	\tilde v(x,p-q)\tilde\psi(x,q)
	=\varpi(p)^2\tilde\psi(x,p),~~~~~x\in\R,~p\in(-k,k).
	\label{sch-eq-nonlocal-tilde}
	\ee
In view of (\ref{v-hat-def}) this is equivalent to the following differential equation in $L^2(-k,k)$. 
	\be
	\left[-\partial_x^2+\widehat v_k(x)\right]\tilde\psi(x)=
	\widehat\varpi^2\tilde\psi(x),~~~~~~x\in\R,
	\label{t-indep-sch-eq}
	\ee
where $\tilde\psi(x)\in L^2(-k,k)$ is given by $\big(\tilde\psi(x)\big)(p):=\tilde\psi(x,p)$, $p\in(-k,k)$, $\widehat\varpi:=\varpi(\widehat p)$, $\varpi(p):=\sqrt{k^2-p^2}$, and $\widehat p:L^2(-k,k)\to L^2(-k,k)$ is the multiplication operator defined by $(\widehat p\,\xi)(p):=p\,\xi(p)$. In particular, 
	\[(\widehat\varpi\,\xi)(p)=\sqrt{k^2-p^2}\,\xi(p), ~~~~~~~p\in(-k,k).\] 
We take the set $C^0(-k,k)$ of continuous functions, $\phi:(-k,k)\to\C$, as the common domain of $\widehat p$ and $\widehat\varpi$, and can easily verify that they are bounded self-adjoint operators. Therefore, we can extend them to everywhere-defined bounded self-adjoint operators acting in $L^2(-k,k)$. It is also easy to see that $\widehat\varpi$ is one-to-one. Its inverse, $\widehat\varpi^{-1}$, is an unbounded self-adjoint operator defined on the range of $\widehat\varpi$;
	\[\dom(\widehat\varpi^{-1})=\sR:=\ran(\widehat\varpi)=\left\{\widehat\varpi\xi~|~\xi\in L^2(-k,k)\;\right\}.\]
Because $\widehat\varpi$ is a one-to-one self-adjoint operator, $\sR$ is a dense subset of $L^2(-k,k)$.

As we noted earlier, the condition that $v$ is a short-range potential implies that the bounded solutions of (\ref{sch-eq-nonlocal}) satisfy (\ref{asym}). We can express this relation in the form:
	\be
	\tilde\psi(x)\to\frac{1}{2\pi}\,\widehat\varpi^{-1}\big(e^{ix\widehat\varpi}A_\pm+
	e^{-ix\widehat\varpi}B_\pm\Big)~~~\for~~~x\to\pm\infty.
	\label{asym2}
	\ee
Clearly,
	\be
	\widehat\varpi^{-1}\big(e^{ix\widehat\varpi}A_\pm+
	e^{-ix\widehat\varpi}B_\pm\Big)=
	\widehat\varpi^{-1}(A_\pm+B_\pm)+
	\widehat\varpi^{-1}\big(e^{ix\widehat\varpi}-\widehat 1)A_\pm+
	\widehat\varpi^{-1}\big(e^{-ix\widehat\varpi}-\widehat 1\big)B_\pm,
	\label{AB-domain-1}
	\ee
where $\widehat 1$ stands for the identity operator acting in $L^2(-k,k)$.
Because $\widehat\varpi^{-1}\big(e^{\pm ix\widehat\varpi}-\widehat 1)$ are everywhere-defined bounded operators, (\ref{asym2}) and (\ref{AB-domain-1}) suggest that 
	\be
	A_\pm+B_\pm\in\sR.
	\label{AB-domain-2}
	\ee
	
Next, let $\Psi:\R\to\sH:=\C^2\otimes L^2(-k,k)$ and $\widehat\bcH(x):\sH\to\sH$ be defined by
	\bea
	&&\Psi(x):=\pi\left[\begin{array}{c}
	e^{-ix\widehat\varpi}
	[\widehat\varpi\tilde\psi(x)-i\partial_x\tilde\psi(x)]\\[3pt]
	e^{ix\widehat\varpi}
	[\widehat\varpi\tilde\psi(x)+i\partial_x\tilde\psi(x)]\end{array}\right],
	\label{Psi-Psi}\\
	&&\widehat\bcH(x):=
	\frac{1}{2}\,e^{-ix\bsigma_3\widehat\varpi}\:\widehat v_k(x)\:\widehat\varpi^{-1}\:\bcK\:e^{ix\bsigma_3\widehat\varpi},
	\label{H-def}
	\eea
where $\tilde\psi:\R\to L^2(-k,k)$ is a given strongly differentiable function with derivative $\partial_x\tilde\psi$. It is easy to see that, according to (\ref{bcK-def}) and (\ref{H-def}), $\widehat\bcH(x)$ is an unbounded operator with (maximal) domain
	\be
	\sD:=\left\{\left.
	\left[\begin{array}{c}
	\phi_+\\
	\phi_-\end{array}\right]\in\sH~\right|~
	e^{ix\widehat\varpi}\phi_++e^{-ix\widehat\varpi}\phi_-\in\sR\right\}.
	\label{domain=}
	\ee
It is also clear from (\ref{Psi-Psi}) and (\ref{domain=}) that
	\be
	\Psi(x)\in\sD.
	\label{Psi-subset}
	\ee	
	
Eq.~(\ref{domain=}) gives the impression that $\sD$ depends on $x$. This is however not true. Because 
	\[e^{ix\widehat\varpi}\phi_++e^{-ix\widehat\varpi}\phi_-=
\phi_++\phi_-+(e^{ix\widehat\varpi}-\widehat 1)\phi_++(e^{-ix\widehat\varpi}-\widehat 1)\phi_-\] and $(e^{\pm ix\widehat\varpi}-\widehat 1)\phi_\pm\in\sR$, Eq.~(\ref{domain=}) implies
	\bea
	\sD&=&\left\{\left.
	\left[\begin{array}{c}
	\phi_+\\
	\phi_-\end{array}\right]\in\sH~\right|~
	\phi_++\phi_-\in\sR\right\}=\sR_-\oplus\sN_0,
	\label{domain=constant}
	\eea
where $\oplus$ marks a direct sum, and
	\begin{align}
	&\sR_-:=\left\{\left.
	\left[\begin{array}{c}
	0\\
	\rho
	\end{array}\right]~\right|~\rho\in\sR\right\}=
	\left\{\left.
	\left[\begin{array}{c}
	0\\
	1\end{array}\right]\widehat\varpi\,\phi~\right|~\phi\in L^2(-k,k)\right\},
	\label{domain=3a}\\
	&\sN_x:=\left\{\left.
	e^{-ix\widehat\varpi\bsigma_3}
	\left[\begin{array}{c}
	1\\
	-1\end{array}\right]\xi~\right|~\xi\in L^2(-k,k)\right\}.
	\label{domain=3b}
	\end{align}
Observe that according to (\ref{AB-domain-2}) and (\ref{domain=constant}),
	\be
	\left[\begin{array}{c}
	A_\pm\\
	B_\pm\end{array}\right]\in\sD.
	\label{AB-domain-3}
	\ee
Another important property of $\sD$ is that it is a dense subset of $\sH$. To see this, we introduce
	\be
	\sR^{2\times 1}:=\left\{\left.
	\left[\begin{array}{c}
	\xi_+\\
	\xi_-\end{array}\right]~\right|~\xi_\pm\in\sR\right\},
	\label{intersection}
	\ee
and use (\ref{domain=constant}) to infer
	\be
	\sR^{2\times 1}\subseteq \sD.
	\label{R2-D}
	\ee
Because $\sR$ is dense in $L^2(-k,k)$, $\sR^{2\times 1}$ is dense in $\sH$. In view of (\ref{R2-D}), this shows that $\sD$ is dense in $\sH$, i.e.,  $\widehat\bcH(x)$ is densely defined.

Next, we obtain an alternative direct-sum decomposition of $\sD$. According to (\ref{domain=}), for every $\left[\begin{array}{c}
	\phi_+\\
	\phi_-\end{array}\right]\in\sD$, there is some $\varsigma\in L^2(-k,k)$ such that $e^{ix\widehat\varpi}\phi_++e^{-ix\widehat\varpi}\phi_-=\widehat\varpi\,\varsigma$. Solving this equation for $\phi_-$ and introducing $\xi:=e^{ix\widehat\varpi}\phi_+$ and $\zeta:=e^{ix\widehat\varpi}\varsigma$, we have $\phi_-=\widehat\varpi\,\zeta-e^{ix\widehat\varpi}\xi$. This calculation shows that 	
	\be
	\left[\begin{array}{c}
	\phi_+\\
	\phi_-\end{array}\right]= 
	\left[\begin{array}{c}
	0\\
	\widehat\varpi\,\zeta\end{array}\right]+
	e^{-ix\widehat\varpi\bsigma_3}
	\left[\begin{array}{c}
	\xi\\
	-\xi\end{array}\right]\:\in\:\sR_-\oplus\sN_x.
	\label{domain=2}
	\ee
Therefore, $\sD\subseteq\sR_-\oplus\sN_x$. It is also easy to see that $\sR_-$ and $\sN_x$ are subspaces of $\sD$. This together with $\sR_-\cap\sN_x=\{0\}$ imply 
	\be
	\sD=\sR_-\oplus\sN_x.
	\label{direct-sum-3}
	\ee
An appealing property of this decomposition is that $\sN_x$ lies in the kernel of $\widehat\bcH(x)$. Therefore, the range of $\widehat\bcH(x)$ coincides with the image of $\sR_-$ under $\widehat\bcH(x)$. It has the following explicit form.
	\be
	\ran[\widehat\bcH(x)]=
	\left\{\left.
	e^{-ix\widehat\varpi\bsigma_3}
	\left[\begin{array}{c}
	\widehat v(x) \zeta\\
	-\widehat v(x) \zeta\end{array}\right]
	~\right|~\zeta\in L^2(-k,k)\right\}.
	\label{range=}
	\ee
In view of (\ref{domain=3b}) and (\ref{range=}),
	\be
	\ran[\widehat\bcH(x)]\subseteq\sN_x\,\subseteq{\rm Ker}[\widehat\bcH(x)]\,\subseteq\sD,
	\label{range-thm}
	\ee
which is consistent with the fact that $\widehat\bcH(x)^2=\widehat\bzero$. The latter follows from (\ref{H-def}) and the fact that $\bcK^2=\bzero$.

The main motivation for the introduction of $\Psi(x)$ and $\widehat\bcH(x)$ is that whenever (\ref{t-indep-sch-eq}) and (\ref{asym2}) holds, they satisfy
	\bea
	&&i\partial_x\Psi(x)=\widehat\bcH(x)\Psi(x),
	\label{t-sch-eq-bcH}\\
	&&\Psi(x)\to \left[\begin{array}{c}
	A_\pm\\
	B_\pm\end{array}\right]~~~\for~~~x\to\pm\infty.
	\label{asym3}
	\eea
In view of (\ref{M-def}), these relations suggest that we identify the fundamental  transfer matrix $\widehat\bM$ for the nonlocal potential $\widehat\sV_k$ with  	
	\be
	\mathop{\mbox{s-lim}}\limits_{{x_\pm\to\pm\infty}} \widehat\bcU(x_+,x_-),
	\label{M=89}
	\ee
where  ``s-lim'' stands for the ``strong limit,'' and $\widehat\bcU(x,x_0)$ is the time-evolution operator for the effective Hamiltonian (\ref{H-def}), i.e., the linear operator $\widehat\bcU(x,x_0):\sH\to\sH$ satisfying
	\be
	i\partial_x\widehat\bcU(x,x_0)=\widehat\bcH(x)\widehat\bcU(x,x_0),~~~~\widehat\bcU(x_0,x_0)=\widehat\bI,
	\label{sch-eq-bcU}
	\ee
where $\widehat\bI$ is the identity operator acting in $\sH$. Notice that (\ref{Psi-subset}) and (\ref{range-thm}) are necessary conditions for making sense of (\ref{t-sch-eq-bcH}) as an equation defined in $\sD$. Because $\widehat\bcH(x)$ is an unbounded operator, we still need to establish the existence of $\widehat\bcU(x,x_0)$ and (\ref{M=89}). In the next section we do this by requiring that $v$ belongs to $\sC_{3,k}$.

\section{Existence of transfer matrix for nonlocal potentials $\widehat\sV_k$}

To establish the existence of the evolution operator $\widehat\bcU(x,x_0)$, we express it as a Dyson series and prove its strong convergence on $\sD$. To achieve this, first we report some preliminary results.

Let $\Phi_0\in\sD$ and $x\in\R$. Then in view of (\ref{domain=3b}) and (\ref{direct-sum-3}), there are unique elements, $\zeta(x)$ and $\xi(x)$, of $L^2(-k,k)$ such that 
	\be
	\Phi_0= 
	\left[\begin{array}{c}
	0\\
	\widehat\varpi\,\zeta(x)\end{array}\right]+
	e^{-ix\widehat\varpi\bsigma_3}
	\left[\begin{array}{c}
	\xi(x)\\
	-\xi(x)\end{array}\right]=
	\left[\begin{array}{c}
	e^{-ix\widehat\varpi}\xi(x)\\
	\widehat\varpi\,\zeta(x)-e^{ix\widehat\varpi}\xi(x)
	\end{array}\right].	
	\label{domain=2n}
	\ee
\noindent {\bf Lemma~3:} Let $\Phi_0\in\sD$ and for all $x\in\R$, $\zeta(x)$ and $\xi(x)$ be the elements of $L^2(-k,k)$ satisfying (\ref{domain=2n}). Then, there are $\fa,\fb\in[0,\infty)$ such that for all $x\in\R$,
	\be
	\parallel\zeta(x)\parallel\leq \fa+\fb\, |x|.
	\label{zeta-bound-main}
	\ee
 Furthermore, the function $\fZ_{\Phi_0}:\R\to[0,\infty)$ defined by 
 	\be
	\fZ_{\Phi_0}(x):=\parallel\zeta(x)\parallel,
	\label{fZ-def}
	\ee 
is continuous.\\[6pt]
 \noindent {\bf Proof:} Let $\zeta_0:=\zeta(0)$ and $\xi_0:=\xi(0)$. Then setting $x=0$ in (\ref{domain=2n}), we have
	\be
	\Phi_0=\left[\begin{array}{c}
	\xi_0\\
	\widehat\varpi\,\zeta_0-\xi_0\end{array}\right].
	\label{Lemma4-eq1}
	\ee
Substituting (\ref{domain=2n}) in the left-hand side of this equation, and solving the resulting equation for $\zeta(x)$ in terms of $\zeta_0$ and $\xi_0$, we find
	\be
	\zeta(x)=\zeta_0+\widehat\varpi^{-1}\left(e^{2ix\widehat\varpi}-\widehat 1\right)\xi_0=
	\zeta_0+2i\widehat\varpi^{-1}\sin(x\widehat\varpi)e^{ix\widehat\varpi}\xi_0.
	\label{zeta-bound}
	\ee
Let $\fa:=\parallel\zeta_0\parallel$ and $\fb:=2\parallel\xi_0\parallel$. Then, (\ref{zeta-bound}) implies
	\be
	\parallel\zeta(x)\parallel
	\;\leq\;\parallel\zeta_0\parallel+2\parallel
	\widehat\varpi^{-1}\sin(x\widehat\varpi)\parallel_0\;
	\parallel e^{ix\widehat\varpi}\xi_0\parallel
	\;\leq\; \fa+\fb\,|x|,
	\nn
	\ee
where we have made use of the fact that $\widehat\varpi^{-1}\!\sin(x\widehat\varpi)$ is a bounded operator, $\parallel
	\widehat\varpi^{-1}\!\sin(x\widehat\varpi)\parallel_0\,\leq |x|$, and $e^{ix\widehat\varpi}$ is a unitary operator. Next, let $x_1,x_2\in\R$. Then, we can use (\ref{zeta-bound}) to show that
	\bea
	\big|\fZ_{\Phi_0}(x_1)-\fZ_{\Phi_0}(x_2)\big|&=&
	\big|\parallel\zeta(x_1)\parallel-\parallel\zeta(x_2)\parallel\big|\nn\\
	&\leq&
	\parallel\zeta(x_1)-\zeta(x_1)\parallel=
	\parallel\widehat\varpi^{-1}\left(e^{2ix_1\widehat\varpi}-e^{2ix_2\widehat\varpi}\right)\xi_0\parallel\nn\\
	&\leq&\parallel2i\widehat\varpi^{-1}
	\sin[(x_1-x_2)\widehat\varpi]e^{i(x_1+x_2)\widehat\varpi}\xi_0\parallel\nn\\
	&\leq& 2 \parallel \widehat\varpi^{-1}\sin[(x_1-x_2)\widehat\varpi] \parallel_0\,
	\parallel e^{i(x_1+x_2)\widehat\varpi} \xi_0\parallel\nn\\
	&\leq& \fb\,|x_1-x_2|.
	\label{continue1}
	\eea
For all $\epsilon\in\R^+$ let $\delta:=1$ if $\fb=0$, and $\delta:=\epsilon/\fb$ if $\fb\neq 0$. Then $\delta\in\R^+$, and in view of (\ref{continue1}), $|x_1-x_2|<\delta$ implies
$\big|\fZ_{\Phi_0}(x_1)-\fZ_{\Phi_0}(x_2)\big|<\fb\,\delta\leq\epsilon$. Hence $\fZ_{\Phi_0}$ is continous.~~~$\square$\vspace{6pt}
	
\noindent {\bf Lemma~4}: Let $\bcK$, $\widehat v_k(x)$, and $\widehat\bcH(x)$ be respectively defined by (\ref{bcK-def}), (\ref{v-hat-def}), and (\ref{H-def}), $\bI$ be the $2\times 2$ identity matrix, $\widehat \bv_k(x):=\widehat v_k(x)\bI$, $\widehat\bvarpi^{-1}:=\widehat\varpi^{-1}\bI$,
	\be
	\widehat\bcL(x):=e^{-ix\widehat\varpi\bsigma_3}
	\left[\begin{array}{cc}
	0 & 0 \\
	1 & 1\end{array}\right]e^{ix\widehat\varpi\bsigma_3}=
	\left[\begin{array}{cc}
	0 & 0 \\
	e^{2ix\widehat\varpi} & 1\end{array}\right],
	\label{bcL=}
	\ee
$n$ be an integer such that $n\geq 2$, $x_1,x_2,\cdots,x_n$ be real numbers, for each $m\in\{1,2,\cdots,n-1\}$, 
	\begin{align}
	\widehat\bs_{m}:=i\widehat\varpi^{-1}
	\sin[(x_{m+1}-x_m)\widehat\varpi]\,\widehat \bv_k(x_m),
	\label{sm-def}
	\end{align}
and 
	\be 
	\widehat\bcB(x_n,x_{n-1},\cdots,x_1):=
	\frac{1}{2}\, e^{-ix_n\widehat\varpi\bsigma_3}
	\widehat\bv_k(x_n)\,
	\widehat\bs_{n-1}\widehat\bs_{n-2}\cdots\widehat\bs_{1}\,\bcK\,
	e^{ix_1\widehat\varpi\bsigma_3}.	
	\label{bcB=}
	\ee
Then $\widehat\bcB(x_n,x_{n-1},\cdots,x_1)$ is a Hilbert-Schmidt operator acting in $\sH$, and 
	\be
	\widehat\bcH(x_n)\widehat\bcH(x_{n-1})\cdots\widehat\bcH(x_1)=\widehat\bcB(x_n,x_{n-1},\cdots,x_1)\,\widehat\bvarpi^{-1}\widehat\bcL(x_1).
	\label{lemma2}
	\ee
\noindent {\bf Proof:} First, we recall that, according to Theorem~2, $\widehat v_k(x)$ and consequently $\widehat\bv_k(x)$ are Hilbert-Schmidt operators. It is also clear that $\widehat\varpi^{-1}\sin[(x_{m+1}-x_m)\widehat\varpi]$ is a bounded operator. Because products of Hilbert-Schmidt and bounded operators are Hilbert-Schmidt, $\widehat\bs_{m}$ and $\widehat\bcB(x_n,x_{n-1},\cdots,x_1)$ are Hilbert-Schmidt operator. To prove (\ref{lemma2}), first we use (\ref{bcK-def}),  (\ref{Pauli}), (\ref{H-def}), and (\ref{bcL=}) to establish the identities:
	\begin{align}
	&\bcK\, e^{i(x_{m+1}-x_{m})\widehat\varpi\bsigma_3}\bcK=
	2i\sin[(x_{m+1}-x_{m})\widehat\varpi]\bcK,
	\label{id1-lemma2}\\
	&\widehat\bcH(x_1)=\widehat\bcH(x_1)\,\widehat\bcL(x_1)=
	\frac{1}{2}\,e^{-ix_1\bsigma_3\widehat\varpi}\:\bcK\:\widehat \bv_k(x_1)\:e^{ix_1\bsigma_3\widehat\varpi}\widehat\bvarpi^{-1}\widehat\bcL(x_1).
	\label{id2-lemma2}
	\end{align}
Substituting (\ref{H-def}) in the left-hand side of (\ref{lemma2}) to obtain the explicit form of $\widehat\bcH(x_n)\widehat\bcH(x_{n-1}),\cdots,\widehat\bcH(x_{1})$ and using (\ref{id2-lemma2}), we find an expression involving terms of the form $\bcK e^{i(x_{m+1}-x_{m})\widehat\varpi\bsigma_3}\bcK$. Eq.~(\ref{lemma2}) follows from this expression and (\ref{id1-lemma2}).~~~$\square$\\[6pt]
\noindent {\bf Lemma~5:} Let $n\in\Z^+$, $\widehat\bcB(x_n,x_{n-1},\cdots,x_1)$ be the operator defined by (\ref{bcB=}) for $n\geq 2$, and
	\be
	\widehat\bcB(x_1):=\widehat\bcH(x_1)\widehat\bvarpi=
	\frac{1}{2}\,e^{-ix\widehat\varpi\bsigma_3}\:\widehat\bv_k(x)\:\bcK\:e^{ix\widehat\varpi\bsigma_3}.
	\label{bcB1}
	\ee
Then,
	\begin{align}
	&\parallel \widehat\bcB(x_1)\parallel_0~\leq~\parallel\widehat v_k(x_1)\parallel_0,
	\label{lemma3-1}\\
	&\parallel \widehat\bcB(x_n,x_{n-1},\cdots,x_1)\parallel_0~\leq~
	\parallel\widehat v_k(x_n)\parallel_0
	\prod_{m=1}^{n-1}|x_{m+1}-x_m|
	\parallel\widehat v_k(x_m)\parallel_0~~~\for~~n\geq 2.
	\label{lemma3}
	\end{align}
{\bf Proof:} In (\ref{bcB=}) and (\ref{bcB1}), we can identify $\bcK$ with $\bcK\widehat\bI$ which is a bounded operator acting on $\sH$ with operator norm $2$. We also know that because $e^{\pm i x\widehat\varpi\bsigma_3}$ are unitary operators, they have unit operator norm. Making use of these observations, Eqs.~(\ref{bcB=}) and (\ref{bcB1}), and $\parallel\widehat\bv_k(x)\parallel_0=\parallel\widehat v_k(x)\parallel_0$, we are led to (\ref{lemma3-1}) and 
	\bea
	\parallel \widehat\bcB(x_n,x_{n-1},\cdots,x_1)\parallel_0&\leq&
	\parallel\widehat v_k(x_n)\parallel_0\;
	\parallel \widehat\bs_{n-1}\parallel_0\;
	\parallel\widehat\bs_{n-2}\parallel_0\cdots
	\parallel\widehat\bs_{1}\parallel_0~~~\for~~n\geq 2.
	\label{lemma3-e1}
	\eea
Furthermore, according to (\ref{sm-def}),
	\bea
	\parallel \widehat\bs_{m}\parallel_0&=&
	\parallel\widehat\varpi^{-1}
	\sin[(x_{m+1}-x_m)\widehat\varpi]\,\widehat\bv_k(x_m)\parallel_0\nn\\
	&\leq&	
	\parallel\widehat\varpi^{-1}
	\sin[(x_{m+1}-x_m)\widehat\varpi]\widehat\bI\parallel_0\,
	\parallel\widehat\bv_k(x_m)\parallel_0\nn\\
	&\leq&|x_{m+1}-x_m|\parallel\widehat v_k(x_m)\parallel_0,
	\label{lemma3-e2}	
	\eea
Relation~(\ref{lemma3}) follows from (\ref{lemma3-e1}) and (\ref{lemma3-e2}).~~~$\square$\\[6pt]
\noindent {\bf Theorem~3:} Let $\widehat\bcH(x)$ be defined by (\ref{H-def}), and $(x_0,x)\in\R^2$ such that $x_0\leq x$. Then the Dyson series, 
	\be
	\widehat\bI+\sum_{n=1}^\infty (-i)^n
         \int_{x_0}^{x} \!\!dx_n\int_{x_0}^{x_n} \!\!dx_{n-1}
         \cdots\int_{x_0}^{x_2} \!\!dx_1\,
         \widehat{\bcH}(x_n)\widehat{\bcH}(x_{n-1})\cdots\widehat{\bcH}(x_1),
         \label{Dyson-bcU}
         \ee
converges strongly to a linear operator $\widehat\bcU(x,x_0)$ defined on $\sD$.
In particular, (\ref{sch-eq-bcU}) has a solution.\\[6pt]
{\bf Proof:} Let $\Phi_0$ be an arbitrary element of $\sD$, and for each $n\in\Z^+$,
	\be
	\Phi_n(x,x_0):= (-i)^n\int_{x_0}^{x} \!\!dx_n\int_{x_0}^{x_n} \!\!dx_{n-1}
         \cdots\int_{x_0}^{x_2} \!\!dx_1\,
         \widehat{\bcH}(x_n)\widehat{\bcH}(x_{n-1})\cdots\widehat{\bcH}(x_1)\Phi_0.
         \label{thm3-eq1}
	\ee
To prove the theorem it suffices to show that the series $\sum_{n=0}^\infty\Phi_n(x,x_0)$ converges. To do this, we introduce
	\be
	\Omega(x_1):=\widehat\bvarpi^{-1}\widehat\bcL(x_1)\,\Phi_0,
	\label{Omega=}
	\ee
and substitute $x_1$ for $x$ in (\ref{domain=2n}) to express $\Phi_0$ in the form
	\be
	\Phi_0=
	\left[\begin{array}{c}
	0\\
	\widehat\varpi\,\zeta(x_1)\end{array}\right]+
	e^{-ix_1\widehat\varpi\bsigma_3}
	\left[\begin{array}{c}
	\xi(x_1)\\
	-\xi(x_1)\end{array}\right].
	\label{domain=2nn}
	\ee
In view of (\ref{bcL=}), (\ref{lemma2}), (\ref{Omega=}), and (\ref{domain=2nn}),
	\begin{align}
	&\Omega(x_1)=
	\widehat\bvarpi^{-1}\left[\begin{array}{c} 0 \\ 
	\widehat\varpi\,\zeta(x_1) \end{array}\right]=\left[\begin{array}{c} 0 \\ 
	\zeta(x_1) \end{array}\right],
	\label{Omega=2}\\
	&\widehat\bcH(x_n)\widehat\bcH(x_{n-1})\cdots\widehat\bcH(x_1)\Phi_0=\widehat\bcB(x_n,x_{n-1},\cdots,x_1)\,\Omega(x_1).
	\label{thm3-eq2}
	\end{align}
Substituting the latter equation in (\ref{thm3-eq1}), we have
	\begin{align}
         \parallel \Phi_n(x,x_0)\parallel\,=&
         \parallel \int_{x_0}^{x} \!\!dx_n\int_{x_0}^{x_n} \!\!dx_{n-1}
         \cdots\int_{x_0}^{x_2} \!\!dx_1\,
         \bcB(x_n,x_{n-1},\cdots,x_1)\Omega(x_1)\parallel\nn\\
         \leq &\int_{x_0}^{x} \!\!dx_n\int_{x_0}^{x_n} \!\!dx_{n-1}
         \cdots\int_{x_0}^{x_2} \!\!dx_1\,
         \parallel \bcB(x_n,x_{n-1},\cdots,x_1)\parallel_0\;
         \parallel \Omega(x_1)\parallel\nn
         \end{align}
In view of (\ref{lemma3-1}), (\ref{lemma3}), and (\ref{Omega=2}), and the fact that $x_{m+1}\geq x_m$, this relation implies
	\be
	\parallel \Phi_1(x,x_0)\parallel\,\leq\,
	\int_{x_0}^{x} \!\!dx_1
	\parallel\widehat v_k(x_1)\parallel_0\;\parallel \zeta(x_1)\parallel,
	\label{lemma3-e11-1}
	\ee
and, for $n\geq 2$,
	\begin{align}
         \parallel \Phi_n(x,x_0)\parallel\,\leq &\int_{x_0}^{x} \!\!dx_n
         \int_{x_0}^{x_n} \!\!dx_{n-1}
         \cdots\int_{x_0}^{x_2} \!\!dx_1 \parallel\widehat v_k(x_n)\parallel_0
	\prod_{m=1}^{n-1}(x_{m+1}-x_m)\!
	\parallel\widehat v_k(x_m)\parallel_0
	\;\parallel \zeta(x_1)\parallel.
	\label{lemma3-e11}
         \end{align}
Next, we observe that, whenever $x_0\leq x_1\leq x_2\leq x$,
	\bea
	\int_{x_0}^{x_2} dx_1 (x_2-x_1) \parallel\widehat v_k(x_1)\parallel_0
	\parallel \zeta(x_1)\parallel
	&\leq&\int_{x_0}^{x_2} dx_1 (x_2-x_0) \parallel\widehat v_k(x_1)\parallel_0
	\parallel \zeta(x_1)\parallel\nn\\
	&\leq&(x_2-x_0)\, F_+(x_2,x_0)\,\leq (x_2-x_0)\, F_+(x,x_0),
	\label{lemma3-e12}
	\eea
where for all $(u,u_0)\in\R^2$,
	\be
	F_+(u,u_0):=\int_{u_0}^{u} dx' \parallel\widehat v_k(x')\parallel_0\,
	\parallel \zeta(x')\parallel_0=
	\int_{u_0}^{u} dx' \parallel\widehat v_k(x')\parallel_0\,\fZ_{\Phi_0}(x'),
	\label{F-def}
	\ee
and we have made use of (\ref{fZ-def}). Notice also that (\ref{lemma3-e11-1}) and (\ref{F-def}) imply
	\be
	\parallel \Phi_1(x,x_0)\parallel\,\leq\,F_+(x,x_0).
	\label{lemma3-e11-2}
	\ee
Lemmas~1 and 4 state that $\parallel\widehat v_k(\cdot)\parallel_0$ and $\fZ_{\Phi_0}$ are respectively piecewise continuous and continuous functions. This implies that the function 
$F_+:\R^2\to\R$ defined by (\ref{F-def}) is continuous. Next, we use (\ref{lemma3-e11}) and (\ref{lemma3-e12}) to show that, for $n\geq 2$,
	\bea
         \parallel \Phi_n(x,x_0)\parallel&\leq&
         F_+(x,x_0)
         \int_{x_0}^{x} \!\!dx_n
         \int_{x_0}^{x_n} \!\!dx_{n-1}
         \cdots\int_{x_0}^{x_3} \!\!dx_2\, \Big[(x_2-x_0)
         \parallel\widehat v_k(x_{2})\parallel_0\times\nn\\
	&&\prod_{m=2}^{n-1}|x_{m+1}-x_m|
	\parallel\widehat v_k(x_{m+1})\parallel_0\Big]\nn\\
	&\leq&F_+(x,x_0)
         \int_{x_0}^{x} \!\!dx_n
         \int_{x_0}^{x_n} \!\!dx_{n-1}
         \cdots\int_{x_0}^{x_3} \!\!dx_2\, \prod_{m=1}^{n-1}|x_{m+1}-x_0|
	\parallel\widehat v_k(x_{m+1})\parallel_0\nn\\
	&\leq&
	\frac{F_+(x,x_0)\, G(x,x_0)^{n-1}}{(n-1)!},
	\label{thm3-expand}
	\eea
where for all $(u,u_0)\in\R^2$,
	\bea
	G(u,u_0)&:=&\int_{u_0}^u dx'\: |x'-u_0|
	\parallel\widehat v_k(x')\parallel_0.
	\label{G-def}
	\eea
Again, because $\parallel\widehat v_k(\cdot)\parallel_0$ is piecewise continuous,
(\ref{G-def}) defines a continuous function $G:\R^2\to\R$ on $\R^2$. It is clear from (\ref{lemma3-e11-2}) and (\ref{thm3-expand}) that, for all $N\in\Z^+$, 
	\bea
	\sum_{n=0}^N\parallel\Phi_n(x,x_0)\parallel 
	&\leq&\parallel\Phi_0\parallel+F_+(x,x_0)
	\sum_{n=1}^N \frac{G(x,x_0)^{n-1}}{(n-1)!}\nn\\[3pt]
         &\leq&\parallel\Phi_0\parallel+\,F_+(x,x_0)\;\Big[e^{\,G(x,x_0)}-1\Big].
         \label{thm3-eq3}
	\eea
This shows that $\sum_{n=0}^\infty\Phi_n(x,x_0)$ converges absolutely. Therefore, it converges. We identify $\widehat\bcU(x,x_0)$ with the operator defined on $\sD$ according to $\widehat\bcU(x,x_0)\Phi_0:=\sum_{n=0}^\infty\Phi_n(x,x_0)$.~~~$\square$\vspace{6pt}

Having established the existence of $\widehat\bcU(x,x_0)$, we address the existence problem for its strong limit as $x_0\to-\infty$ and $x\to+\infty$, which we identify with the transfer matrix $\widehat\bM$.\\[6pt]		
\noindent {\bf Theorem~4:} Let $k\in\R^+$ and $v\in\sC_{3,k}$. Then $\mathop{\mbox{s-lim}}\limits_{{x_\pm\to\pm\infty}} \widehat\bcU(x_+,x_-)$ exists as a linear operator defined on $\sD$.\\[6pt]
{\bf Proof:} First, we recall that because $v\in\sC_{3,k}$, there are $\alpha,\beta,\sigma\in\R^+$ such that $\sigma>3$ and $\widehat v_k(\cdot)$ satisfies (\ref{lemma1}). We also note that using the Dyson series expansion (\ref{Dyson-bcU}) of $\widehat\bcU(x_,x_0)$ and its strong absolute convergence we can prove the following identity \cite{Reed-Simon2}.
	\be
	\widehat\bcU(x_3,x_2)\,\widehat\bcU(x_2,x_1)=\widehat\bcU(x_3,x_1)~~~
	\for~~~x_1\leq x_2\leq x_3.
	\label{composition}
	\ee
Repeated use of this relation, we can express $\widehat\bcU(x_+,x_-)$ in the form
	\be
	\widehat\bcU(x_+,x_-)=
\widehat\bcU(x_+,\alpha)\,\widehat\bcU(\alpha,-\alpha)\,
\widehat\bcU(-\alpha,x_-)~~~\for~~~\pm x_\pm\geq\alpha,
	\label{composition}
	\ee
which reduces the existence problem for 
	$\mathop{\mbox{s-lim}}\limits_{{x_\pm\to\pm\infty}} \widehat\bcU(x_+,x_-)$ to that for $\mathop{\mbox{s-lim}}\limits_{{x_+\to+\infty}} \widehat\bcU(x_+,\alpha)$ and \linebreak $\mathop{\mbox{s-lim}}\limits_{{x_-\to-\infty}} \widehat\bcU(-\alpha,x_-)$. To prove the existence of $\mathop{\mbox{s-lim}}\limits_{{x_+\to+\infty}} \widehat\bcU(x_+,\alpha)$, it is sufficient to show that as $x\to+\infty$ the right-hand side of (\ref{thm3-eq3}) with $x_0=\alpha$ remains bounded by a positive real number. We show this by proving the existence of $\gamma ,\delta \in\R^+$ such that for all $x_+\geq\alpha$, 
	\begin{align}
	&F_+(x_+,\alpha)\leq \gamma , &&G(x_+,\alpha)\leq \delta.
	\label{FG-bound}
	\end{align}
According to Lemma~3, there are $\fa,\fb\in[0,\infty)$ such that for all $x\in\R$, $\fZ_{\Phi_0}(x)\leq\fa+\fb|x|$. Using this in (\ref{F-def}), we have
	\be
	F_+(x_+,\alpha)\leq \fa\, f_0(x_+,\alpha)+\fb\, f_1(x_+,\alpha),
	\label{Fff-bound}
	\ee
where
	\be
	 f_\ell(x,x_0):=\int_{x_0}^{x} dx'\; |x|^\ell
	\parallel\widehat v_k(x')\parallel_0,\quad\quad\quad\ell\in\{0,1\}.
	\label{f-ell}
	\ee	
Next, we use (\ref{G-def}) to infer that whenever $x_+\geq\alpha$,
	\be
	G(x_+,\alpha)=\int_{\alpha}^{x_+} dx'\,(x'-\alpha)
	\parallel\widehat v_k(x')\parallel_0\;\leq\,
	\int_{\alpha}^{x_+} dx'\,x'
	\parallel\widehat v_k(x')\parallel_0=f_1(x_+,\alpha).
	\label{G-bound1}
	\ee
Relations (\ref{Fff-bound}) -- (\ref{G-bound1}) reduce the proof of the existence of $\gamma$ and $\delta$ to finding upper bounds
on $f_\ell(x_+,\alpha)$. In view of (\ref{lemma1}) and (\ref{f-ell}), and the fact that $x_+\geq\alpha>0$, $\sigma>3$, and $\ell\in\{0,1\}$,
	\bea
	f_\ell(x_+,\alpha)&=&
	\int_{\alpha}^{x_+} dx'\;|x'|^\ell
	\parallel\widehat v_k(x')\parallel_0
	\;\leq 2\pi\beta\int_{\alpha}^{x_+} dx'\: 
	\frac{|x'|^\ell}{(1+|x'|)^\sigma}\nn\\
	&\leq& 2\pi\beta\int_{\alpha}^{x_+} dx'\: x'^{\ell-\sigma}\:=\:
	\frac{2\pi\beta}{\sigma-\ell-1}\left(\frac{1}{\alpha^{\sigma-\ell-1}}-
	\frac{1}{x_+^{\sigma-\ell-1}}\right)
	<\frac{2\pi\beta}{(\sigma-2)\,\alpha^{\sigma-\ell-1}}.~~~~
	\label{thm3-last}
	\eea
In view of (\ref{Fff-bound}), (\ref{G-bound1}), and (\ref{thm3-last}), the following choices for $\eta$, $\gamma $, and $\delta $ satisfy (\ref{FG-bound}).
	\begin{align}
	&\gamma :=\frac{2\pi\beta(\fa+\fb\,\alpha)}{(\sigma-2)\alpha^{\sigma-1}},
	&&\delta :=\frac{2\pi\beta}{(\sigma-2)\alpha^{\sigma-2}}.
	\nn
	\end{align}
This observation together with (\ref{thm3-eq3}) show that $\sum_{n=0}^\infty\parallel\Phi_n(x_+,\alpha)\parallel<\infty$ which implies the existence of 
$\mathop{\mbox{lim}}\limits_{{x_+\to+\infty}} \sum_{n=0}^\infty \Phi_n(x_+,\alpha) $ for every choice of $\Phi_0\in\sD$. Hence $\mathop{\mbox{s-lim}}\limits_{{x_+\to+\infty}} \widehat\bcU(x_+,\alpha)$ exists as a linear operator defined on $\sD$. To prove the existence of $\mathop{\mbox{s-lim}}\limits_{{x_-\to-\infty}} \widehat\bcU(-\alpha,x_-)$, we first use (\ref{lemma3-e11}) to establish the following inequality for $n\geq 2$.
	\bea
         \parallel \Phi_n(x,x_0)\parallel
         &\leq&\frac{f_0(x,x_0)F_-(x,x_0)[-G(x_0,x)]^{n-2}}{(n-2)!},
         \label{lemma3-e11-m}
         \eea
where $f_0$, and $G$ are respectively defined by (\ref{f-ell}) and (\ref{G-def}), for all $(u,u_0)\in\R^2$,
	\be
	F_-(u,u_0):=\int_{u_0}^{u} \!\!dx' |u-x'|\parallel\widehat v_k(x')\parallel_0\;\parallel \zeta(x')\parallel,
	\label{Fm=def}
	\ee
and we have made use of the identity,
	\be
	\int_{x_0}^x dx'(x-x')\parallel\widehat v_k(x')\parallel_0=
	-\int_{x}^{x_0} dx'|x-x'|\parallel\widehat v_k(x')\parallel_0=-G(x_0,x),
	\label{G-Id}
	\ee
which follows from (\ref{G-def}). Notice that similarly to $f_0(x,x_0)$, $F_-(x,x_0)$ and $-G(x_0,x)$ take nonnegative real values for $x\geq x_0$. According to (\ref{lemma3-e11-2}) and (\ref{lemma3-e11-m}),
	\bea
	\sum_{n=0}^N\parallel\Phi_n(x,x_0)\parallel 
	&\leq&\parallel\Phi_0\parallel+\parallel\Phi_1(x,x_0)\parallel+
	\,f_0(x,x_0)F_-(x,x_0)
	\sum_{n=2}^N \frac{[-G(x_0,x)]^{n-2}}{(n-2)!}\nn\\[3pt]
         &\leq&\parallel\Phi_0\parallel+F_+(x,x_0)+
         f_{0}(x,x_0)F_-(x,x_0)\; e^{-G(x_0,x)} .
         \label{thm3-eq3m}
	\eea
Next, we use (\ref{zeta-bound-main}), (\ref{F-def}), and (\ref{f-ell}) to show that 
	\be
	0\leq F_+(-\alpha,x_-)\leq \fa\, f_0(-\alpha,x_-)+\fb\, f_1(-\alpha,x_-)~~~\for~~~x_-\leq-\alpha.
	\label{Fff-bound-m}
	\ee
Repeating the calculations leading to (\ref{thm3-last}) with $(x_+,\alpha)$ replaced with $(-\alpha,x_-)$, we obtain
	\bea
	0\leq f_\ell(-\alpha,x_-)<\frac{2\pi\beta}{(\sigma-2)\,\alpha^{\sigma-\ell-1}}~~~\for~~~x_-\leq-\alpha.
	\label{thm4-last}
	\eea
Furthermore, we can use (\ref{lemma1}), (\ref{zeta-bound-main}), (\ref{f-ell}), (\ref{Fm=def}), (\ref{G-Id}), and the fact that $x_-\leq-\alpha<0$, and $\sigma>3$,
to establish the following relations.
	\bea
	0\leq F_-(-\alpha,x_-)&=&\int_{x_-}^{-\alpha} \!\!dx' (-x'-\alpha)\parallel\widehat v_k(x')\parallel_0\;\parallel \zeta(x')\parallel
	\;\leq\;\int_{x_-}^{-\alpha} \!\!dx' (-x')\parallel\widehat v_k(x')\parallel_0\;\parallel \zeta(x')\parallel\nn\\
	&\leq&2\pi\beta\int_{x_-}^{-\alpha} \!\!dx' \frac{(-x')(\fa+\fb|x'|)}{(1+|x'|)^\sigma}\;\leq\;2\pi\beta\int_{x_-}^{-\alpha} \!\!dx' \frac{(-x')(\fa+\fb|x'|)}{|x'|^\sigma}\nn\\
	&\leq&2\pi\beta\int_{x_-}^{-\alpha} \!\!dx' \left[
	\fa(-x')^{1-\sigma}+\fb(-x')^{2-\sigma}\right]	\;\leq\;
	\frac{2\pi\beta(\fa+\fb\,\alpha)}{(\sigma-3)\alpha^{\sigma-2}},
	\label{Fm-bound}\\
	0\leq -G(-\alpha,x_-)&=&\int_{x_-}^{-\alpha} \!\!dx' (-x'-\alpha)\parallel\widehat v_k(x')\parallel_0\;\leq\;\int_{x_-}^{-\alpha} \!\!dx' (-x')\parallel\widehat v_k(x')\parallel_0\nn\\
	&\leq&2\pi\beta\int_{x_-}^{-\alpha} \!\!dx' \frac{-x'}{(1+|x'|)^\sigma}\;\leq\;2\pi\beta\int_{x_-}^{-\alpha} \!\!dx' (-x')^{\sigma-1}\;\leq\;
	\frac{2\pi\beta}{(\sigma-2)\alpha^{\sigma-2}}.
	\label{Gm-bound}
	\eea
In view of (\ref{thm3-eq3m}) -- (\ref{Gm-bound}),  $\mathop{\mbox{lim}}\limits_{{x_-\to-\infty}} \sum_{n=0}^\infty \Phi_n(-\alpha,x_-) $ exists for every choice of $\Phi_0\in\sD$. This implies the existence of $\mathop{\mbox{s-lim}}\limits_{{x_-\to-\infty}} \widehat\bcU(-\alpha,x_-)$ on $\sD$.~~~$\square$\vspace{12pt}

\noindent{\bf Acknowledgements:}
We wish to express our gratitude to Varga Kalantarov for helpful discussions. This work has been supported by the Scientific and Technological Research Council of Turkey (T\"UB\.{I}TAK) in the framework of the project 120F061 and by Turkish Academy of Sciences (T\"UBA).\vspace{12pt}


\section*{Appendix: Scattering amplitude and transfer matrix in 2D}

When $v$ is a short-range potential, the asymptotic expression for the scattering solutions of (\ref{sch-eq}) may be put in the form
	\[\psi(\bfr)\to
	\frac{1}{2\pi}\Big[e^{i\bk_0\cdot\bfr}+\sqrt{\frac{i}{kr}}\,e^{ikr}\ff(\theta_0,\theta) \Big]~~~\for~~~r\to\infty,\]
where $\bk_0\in\R^2$ is the incident wave vector, $\bfr:=(x,y)$,  $(r,\theta)$ are the polar coordinates of $\bfr$, $\theta_0$ is the angle between $\bk_0$ and the $x$-axis, and $\ff(\theta_0,\theta)$ is the scattering amplitude of the potential $v$ for the incident wavenumber $k$, \cite{adhikari}. Let $p_0:=k\sin\theta_0$ and $p:=k\sin\theta$. Then $\varpi(p)=k|\cos\theta|$, $\varpi(p_0)=k|\cos\theta_0|$, and we have \cite{pra-2021}:
	\begin{itemize}
	\item[1)] For $\theta_0\in(-\frac{\pi}{2},\frac{\pi}{2})$: $A_-(p)=2\pi\varpi(p_0)\delta(p-p_0)=2\pi\delta(\theta-\theta_0)$, $B_+(p)=0$, 
	\[\ff(\theta_0,\theta)=-\frac{i}{\sqrt{2\pi}}\times \left\{
	\begin{array}{ccc}
	A_+(k\sin\theta)-2\pi\delta(\theta-\theta_0) &\for & \theta\in(-\frac{\pi}{2},\frac{\pi}{2}),\\[6pt]
	B_-(k \sin\theta) &\for & \theta\in(\frac{\pi}{2},\frac{3\pi}{2}),\end{array}\right.\]
	$A_+= 2\pi \widehat M_{11}\:\delta_{p_0}+\widehat M_{12}B_-$,  and $B_-$ satisfies $\widehat M_{22}\,B_-=-2\pi\widehat M_{21}\delta_{p_0}$, where $\delta_{p_0}$ stands for the Dirac delta function centered at $p_0$, i.e., $\delta_{p_0}(p):=\delta(p-p_0)$. 

	\item[2)] For $\theta_0\in(\frac{\pi}{2},\frac{3\pi}{2})$: $A_-(p)=0$, $B_+(p)=2\pi\varpi(p_0)\delta(p-p_0)=2\pi\delta(\theta-\theta_0)$, 
	\[\ff(\theta_0,\theta)=\frac{i}{\sqrt{2\pi}}\times \left\{
	\begin{array}{ccc}
	A_+(k \sin\theta) &\for & \theta\in(-\frac{\pi}{2},\frac{\pi}{2}),\\[6pt]
	B_-(k \sin\theta)-2\pi\delta(\theta-\theta_0) &\for & \theta\in(\frac{\pi}{2},\frac{3\pi}{2}),\end{array}\right.\]	
	$A_+=\widehat M_{12}B_-$, and $B_-$ solves
	$\widehat M_{22}\,B_-=2\pi \delta_{p_0}$.

	\end{itemize}
In particular, given the transfer matrix $\widehat\bM$, the determination of the scattering amplitude $\ff(\theta_0,\theta)$ requires the solution of $\widehat M_{22}\,B_-=-2\pi\widehat M_{21}\delta_{p_0}$ and $\widehat M_{22}\,B_-=2\pi \delta_{p_0}$. The existence of a well-defined scattering problem for the potential is therefore linked with the triviality of the kernel of $\widehat M_{22}$. 

\ed